\documentclass[aps,twocolumn,amsfonts,showpacs]{revtex4}
\usepackage{epsfig,amsopn,amsmath}
\usepackage{multirow,bigdelim}

\usepackage{color}
\usepackage{lipsum}

\begin{document}
	
	\title{A New ($\aleph$) Stochastic Quenched Disorder Model for Interaction of Network-\\
		Master node}
	\author{Anderson A.  Ferreira}\email{ferreira07@unifesp.br}
	\affiliation{%
		Departamento de F\'{\i}sica, Universidade Federal de S\~ao Paulo, \\
		Diadema-SP, Brazil
	}
	
	\author{Leandro A. Ferreira}
	\affiliation{%
		Instituto de Matem\'atica e Estat\'{i}stica, Universidade de S\~ao Paulo, \\
		S\~ao Paulo-SP, Brazil
	}%
	%
	
	\author{Fernando Fagundes Ferreira}\email{ferfff@usp.br}
	\affiliation{%
		Escola de Artes, Ci\^encias e Humanidades, Universidade de S\~ao Paulo,
		03828-000 S\~ao Paulo, Brazil
	}%
	
	%
	

\begin{abstract}
	Human beings live in a networked world in which information spreads very fast thanks to the advances in technology. In the decision processes or opinion formation there are different ideas of what is collectively good but they tend to go against the self interest of a large amount of agents. Here we proposed a simple model who try to mimic, in some extent, this class of problems. We look at interacting agents connected in networks to analyze the multi-agent decision stochastic process. Each individual decision may be influenced by others’ decisions. We consider a first neighbor interaction where agents are spread through a uni-dimensional network. Some agents are also connected to a hub, or master node, who has preferential values (or orientation). The role of master node is to persuade some individuals to follow a specific orientation, subject to a probability of successful persuasion. The connections between master node and the network society are quenched in disorder. Despite its simplicity, we found a phase transition from disorder to order for three different control parameters. We also discuss how this model may be useful as a framework to study the spread of morality, innovation, opinion formation and consensus. Is important to recall the route from disorder to order in social systems still a great challenge. We hope to contribute with a novel approach to model a this issues. 
\end{abstract}

\pacs{05.70.Fh,05.70.Ln,02.50.Ey,89.75.Fb,89.75.Hc}

\maketitle
\section*{Introduction}

Failure in cooperating can threaten existence itself.  Conflicts and issues such as wars, corruption, loss of liberty and tyrannies, environmental degradation, deforestation, among others pose great problems and are a testament to humanity’s inability of cooperate in a suitable level. These examples show that we still do not have a complete understanding of the mechanism which drives the collective toward to a common goal and hence to avoid the tragedy of the commons \cite{hardin1968tragedy,ostrom2002drama}. Despite that, altruism, cooperation and moral norms still being improved to outcompete behaviors of free riders, selfishness and immoral \cite{axelrod1981emergence,nowak2006five}.

Living organisms and human beings are characterized by autonomy. However, they tend to be prone to selfishness, a bias that may bring harsh damage to their survival as well as the environment, maybe due to ambitions and potential short-sightedness. The assumption that living organisms are selfish has been accepted by many branches of contemporary science. For example, the inclusive fitness theory, in ecology, in which egoism has biological roots \cite{alexander2017biology}. A similar idea arises in neoclassical economic theory, which hypothesized that all choices, no matter if altruistic or self-destructive, are designed to maximize personal utility \cite{harrison1985egoism}. Thus, decisions are motivate by self-interest.

One of the most interesting issues to be addressed in the context of the present study is morality. The human free will combined with its selfish/altruist nature may create a plethora of different patterns over several types of social systems. How does morality emerge? This question probably does not have a simple and unique answer. Many thinkers from ancient times to today, tries to unravel this phenomenon. As a legacy we have a body of theories that seek to understand and explain the emergence of morality in different societies.  Thomas Hobbes \cite{hobbes2006leviathan,cranston1989jean} was one of the first modern philosopher to offer a naturalist principle to ethics. In his theory, ethics emerge when people understand the necessary conditions to live well. According to Hobbes, these conditions are defined by imposition of equality of rights, by means of an absolute Sovereign, due to the necessity of self preservation and by establishing deals among individuals. Latter on, Rousseau  proposed  that life in community can lead to the loss of individual freedom since the subjects must fulfill a social contract expressed through laws and institutions \cite{rousseau2002social}. Unlike the philosophers who attributed to reason the capacity to conceive morality, Durkheim understood it as a result of a set of social interactions and culture elaborated throughout history \cite{durkheim1973emile,marks1974durkheim}. But these are part of a small selection of seminal works about a theme  hallmarked by an intriguing and challenge scientific problem. This debate continues in different areas such as Psychology \cite{haidt2008social,haidt2008morality}, Political Science, Philosophy, Antropology \cite{ostrom2000collective}, Education, Economics and Ecology \cite{huberman1993evolutionary,santos2006evolutionary}.

To further advance the long discussion on morality or cooperation, we need to understand some specific mechanisms of social interaction in various scenarios with different individual degrees of freedom, effective individual choices, and consider that these choices are influenced locally by peers' s opinions \cite{rowe2002nicomachean,rachels1993elements,brandt1996facts}. Different levels of freedom (free choice), control (supervision) and social dynamics  impact the individual capacity to  fulfillment of the social contract and hence should lead to different degrees of morality or cooperation at the collective level.

It is a well-known fact that a system of interacting linked individuals can work together to reach a collective goal.  Understanding how decentralized actions can lead to these results has been a topic of study in the literature for decades. The focus of our study is the role of a master node, connected to some members of a society, may drive the pursued ideal by collectiveness. The topology formed by a master node connected to a network may represent many situations in social systems: law enforcement and citizens \cite{zaklan2009analysing, orviska2003tax,rose2001trust}; moral and community \cite{rowe2002nicomachean,rachels1993elements,brandt1996facts,haidt2008morality};  beliefs and member of churchs \cite{stark1980networks,shi2016evolution,galam1991towards}; cooperation and egoism \cite{ohtsuki2006simple, beersma2003cooperation}; tax evasion and fiscal country, among others. In all examples, individuals do not share the same goals, due to the incentives in acting against the common good.

We approach this issue using a stochastic quenched disorder model to study the consensus formation \cite{castellano2009statistical,gonzaga2018quenched}. In this model, individuals are autonomous to make decisions based on their own opinions or let decisions be influenced by a local social group or/and by the presence of a norm (master) that reinforces preferential behaviors. The individual decisions are binaries (0 or 1) and the collective decision is the average collective decision.

This model belongs to a class of nonequilibrium systems \cite{sampaio2011block, medeiros2006domain,stinchcombe2001stochastic}. We found absorbing states phase transitions with respect to three distinct order parameter \cite{dickman1998self,evans2000phase,hinrichsen2000non,de2015generic}. From a statistical mechanics point of view, phase transition in nonequilibrium sytems are studied by fundamental concept as scaling and universality class \cite{lubeck2004universal,lubeck2003universal,odor2004universality,de2005spontaneous}, which may reserve some unexpected results \cite{hexner2015hyperuniformity,van2002universality}.

The remainder of the paper is organized as it follows: in Sec. \ref{sec:model} we define the model and introduce the general notation. To reduce the number of variables, a parametrization was proposed to help us to analyze the model. In Sec. \ref{sec:results} we explain the simulation and present the outcomes. Analysis and discussions of our results are found in Sec. \ref{sec:discussion}. Finally, in Sec. \ref{sec:conclusion} contains conclusion and an outlook on future work.

\section{The Model}
\label{sec:model}

\begin{figure}[htb]
		\begin{center}
			\includegraphics[width=0.4\textwidth]{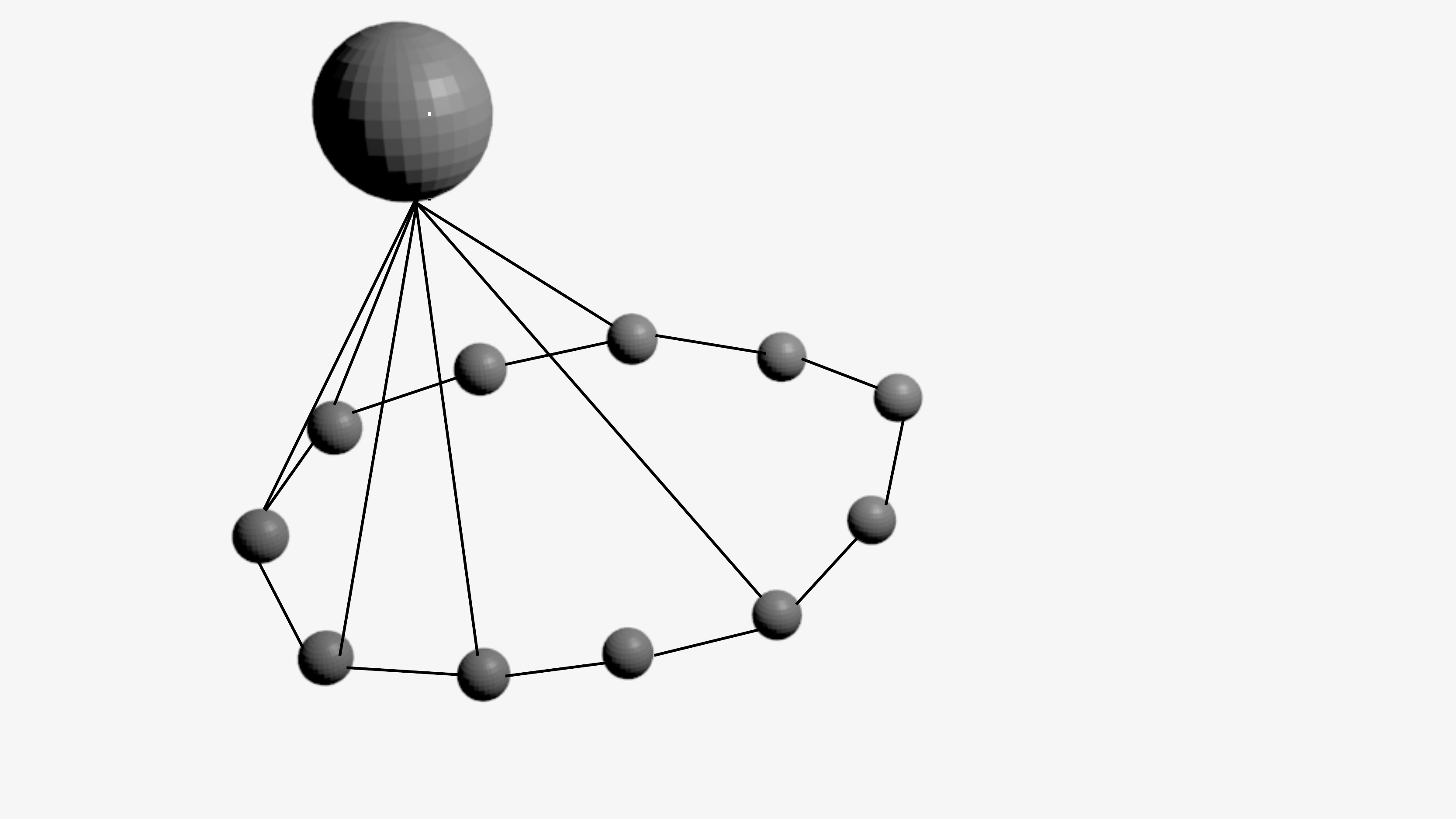}
		\end{center}
		\caption{ Model representation. Small spheres represent interacting particles or agents. Each particle is in the state $s=0,1$ with probability $w_s$.Large sphere respresent the master state, in which the interaction strength with particles is fixed (denoted by $r$) and the links are quenched disorder with density denoted by $\rho$.}
		\label{fig:model}
	\end{figure}

In Figure \ref{fig:model} we illustrate our model. It consists of a ring formed by  nodes with periodic boundary conditions. Initially, each node represents particles or agents which may assume two different states $s=0,1$  with probability $w_s$. If , this means that there may be an intrinsic tendency or preference of particles for a determined state $s$. So, in principle, this is a particle property. They interact with first neighbors (on the left and on the right). Moreover, we introduce a master node illustrated by a large sphere on the top of the ring in Figure \ref{fig:model}. The master node connects with particles located in the ring with probability  in the initial time (quenched disorder). The interaction strength between master and connected agents is denoted by $r$. The general configuration of the system is given by $(\beta_i, \Gamma)$, with $i=1, \dots, L$ representing the individual states and $\Gamma$ representing the existence of a connection between master node and node $i$.

We have two different types of interaction. First of all we identify the interaction between particle $i$ and its neighbors $i-1$ and $i+1$ with the state of particle $i$ dependent on the state of its neighbors $(\beta_{i-1}~ |~ \beta_{i+1}$. In the absence of a master, particle $i$ will align with the majority in the neighborhood, a situation which lead to consensus. If there are differences between neighbor states (frustration), the decision is probabilistic. If the particle $ i $  is in the state $ s $, she/he switches to another  state with probability $ w_s $. When there is a mixed state $(0|1)$,  the probability transition depends on the state  of of the particle $i$.

The second type occurs with presence of the master node $\Gamma_i=1$, which belief or orientation is equal to 1.  The probability of the particles being influenced by the master particle is equal to $ r $. Suppose the particle $i$ is in the state $s = 1$. If most of your neighbors are in the state $s = 0$ then the particle will change to state $s=0$ with probability $ 1-r $. However, if there is frustration between the neighbors $(0 | 1)$ or $(1 | 0)$, then the particle $i$ changes to the state $(s = 0)$ with probability $r_1$. Now suppose that the particle $i$ is in the state $s = 0$. In the case where neighbors are in the state $s=1$, due to peer pressure (majority) and also due to the influence of the master, the particle will change to state $s=1$. There are two conflict situation. When the majority of neighbors are in the state $s=0$, with probability $q$ the particle $i$ will change to state $s=1$ or remain in the same state with probability $p=1-q$. The second situation there is frustration between neighbors. Now, with probability $r_0$ particle $i$ will change to state $s=1$ and stay in the same state with probability $r_1=1-r_0$.

We summarize all the situation describe above in the table \ref{tab1}. The notation used in this table is based on the state value of $\Gamma$ and $\beta_i \in (\beta_1;\beta_2;\ldots;\beta_L)$. In the columns we represent the interaction between particles $[\beta_{i-1}, \beta_{i+1}]$ while we represent the interaction between master and particles $(\beta_i, \Gamma)$ in the lines. In the latter, we have four combinations, two without a master $(1,0)$ and $(0,0)$ where particle $i$ is in the state $s=1$ and $s=0$ respectively. Others two combinations with the presence of master: $(1,1)$ and $(0,1)$ where particle $i$ is in the state $s=1$ and $s=0$ respectively. On the top of the arrows we represent the probability transitions respectively. 


	\begin{table*}[!htb]
		\centering
		\caption{This table summarizes all the possible state transitions in the model}
		\begin{tabular}{|c|p{2.0cm}|p{2.0cm}|p{2.0cm}|p{2.0cm}|}
			\hline
			$(\beta_i,\Gamma)$  & \multicolumn{4}{|c|}{$\big{[}\beta_{i-1}|\beta_{i+1}\big{]}$}\\
			\cline{2-5} & $\big{[}1|1\big{]}$ & $\big{[}0|1\big{]}$ & $\big{[}1|0\big{]}$ & $\big{[}0|0\big{]}$\\
			\hline $(1,1)$   & $1_i\xrightarrow{1}1_i$ & $1_i\xrightarrow{r_1}0_i$ $1_i\xrightarrow{1-r_1}1_i$& $1_i\xrightarrow{r_1}0_i$ $1_i\xrightarrow{1-r_1}1_i$ & 
			$1_i\xrightarrow{r}1_i$ $1_i\xrightarrow{1-r}0_i$ \\
			\hline $(0,1)$   &$0_i\xrightarrow{1}1_i$  & $0_i\xrightarrow{r_0}1_i$ $0_i\xrightarrow{1-r_0}0_i$ & $0_i\xrightarrow{r_0}1_i$ $0_i\xrightarrow{1-r_0}0_i$ & 
			$0_i\xrightarrow{q}1_i$ $0_i\xrightarrow{1-q}0_i$\\
			\hline $(1,0)$   & $1_i\xrightarrow{1}1_i$ & $1_i\xrightarrow{w_1}0_i$ $1_i\xrightarrow{1-w_1}1_i$ & $1_i\xrightarrow{w_1}0_i$ $1_i\xrightarrow{1-w_1}1_i$ & 
			$1_i\xrightarrow{1}0_i$\\
			\hline $(0,0)$   & $0_i\xrightarrow{1}1_i$ & $0_i\xrightarrow{w_0}1_i$ $0_i\xrightarrow{1-w_0}0_i$ & $0_i\xrightarrow{w_0}1_i$ $0_i\xrightarrow{1-w_0}0_i$ & 
			$0_i\xrightarrow{1}0_i$\\
			\hline
		\end{tabular}
		\label{tab1}
	\end{table*}

	Aiming to analyse this model, we propose the following  parametrization described in the subsection below.

	\subsubsection*{Parameterization}
	
	Let us choose some constraints to the parameters $p, q, r_0, r_1, w_0$ and $w_1$ in terms of $r$ ('' influence of master node'') and $\Delta=w_0-w_1$, which is the intrinsic state tendency of agents, and $\rho$. For simplicity, we take $q=r$.  Since $w_1=1-w_0$ we may write

	\begin{equation}
	\Delta=w_0-w_1 =1-2w_1.
	\end{equation}

	The parameter $\Delta$ measures the natural nature of an element or particle be in the state $s=0 (1)$ when $\Delta>0 (<0)$ in the absence of any interaction or influence.
	If $0<w_1<\frac{1}{2} $  the individuals, in average, will behave against the norm or the common good. In this case, $\Delta>0$ which means that the system has a tendency to be opposite to the master (selfish or immoral). We are interested in studying how such a system undergoes to a phase dominated by the main orientation (cooperative or exclusively moral), so we will vary the parameter in the interval $0<\Delta<1$.
	
	
	The probabilities $r_0$ and $r_1$ should be parameterized so that when the master's influence is null ($r=0$) we have $r_0=w_0$ and $r_1=w_1$. Otherwise, when $r=1$  we  should have necessarily $r_0=1$ and $r_1=0$. The simplest way is through a linear parameterization

	\begin{eqnarray}
	&&r_0=r+(1-r)(\frac{1-\Delta}{2}),\\
	&&r_1=(1-r)(\frac{1+\Delta}{2}).
	\end{eqnarray}
	
	\noindent The parametrized version of the model has only three free parameters:
	
	\begin{equation}
	0\leq r \leq 1,;\;\;\;\;0\leq \Delta \leq 1,;\;\;\;\mbox{and}\;\;\;\;\; 0\leq\rho\leq 1.
	\end{equation}
	
	
	\section{Numerical Simulations}
	\label{sec:results}
	
	At each time step, a node $i$ is chosen randomly among the L nodes and the Monte Carlo simulations were performed on a ring with periodic boundaries. The Game Plan ($GP$) takes place:  
	
	{\bf 1} : A node $i$ is chosen randomly between the L nodes, then 
	
	{\it
	\begin{itemize}
		 
	 \item $GP_{A}$: If the site choosen is in the state $(1,1)$, then the dynamics follows according to the first line of table 1;
	 
	 \item $GP_{B}$: If the site choosen is in the state $(0,1)$, then the dynamics follows according to the second line  of table 1; 
	
	\item $GP_{C}$: If the site choosen is in the state $(1,0)$, then the dynamics follows according to the third line of table 1; 
	
	\item $GP_{D}$: If the site choosen is in the state $(0,0)$, then the dynamics follows according to  the fourth line of table 1.

	\end{itemize}
}

	\rm
	
	 {\bf 2} : After any (or not) change in the state of system, the instant $t$ is rasing  by $\Delta t=1/L$.  Any observable $\langle \theta(t) \rangle$ of the model is calculated by performing averages on the different positional configurations of master node (for fixed density  density of links $\rho$) and on the different initial configuratons over individual node  for a given $\Delta$ and $r$ in the period from 0 to $t$. For instance, the global system state is given by the average quantity $\langle\langle M(t)\rangle\rangle \;\;=\;\;\langle\langle\sum_i \beta_i(t)/L\rangle\rangle$.	In the absence a master influence $(r=0)$ the system presents and evolves to two absorbing states $s=0$ and $s=1$.

	When we activated the influence of a master over the elements in the network $(\rho \neq 0)$ the situation changes to an unexpected result. To understand this we shall take into account that $M$ is a function of $t, ~\Delta, ~\rho$ and $r$. The system goes to a phase transition by varying the control parameters. Since it takes place in a 4D space, to emphasize the character of the phase transitions we take three distinct projection in the phase diagram of the model.
	
\begin{figure}[htb]
		\begin{center}
					\includegraphics[height=8cm, width=8cm]{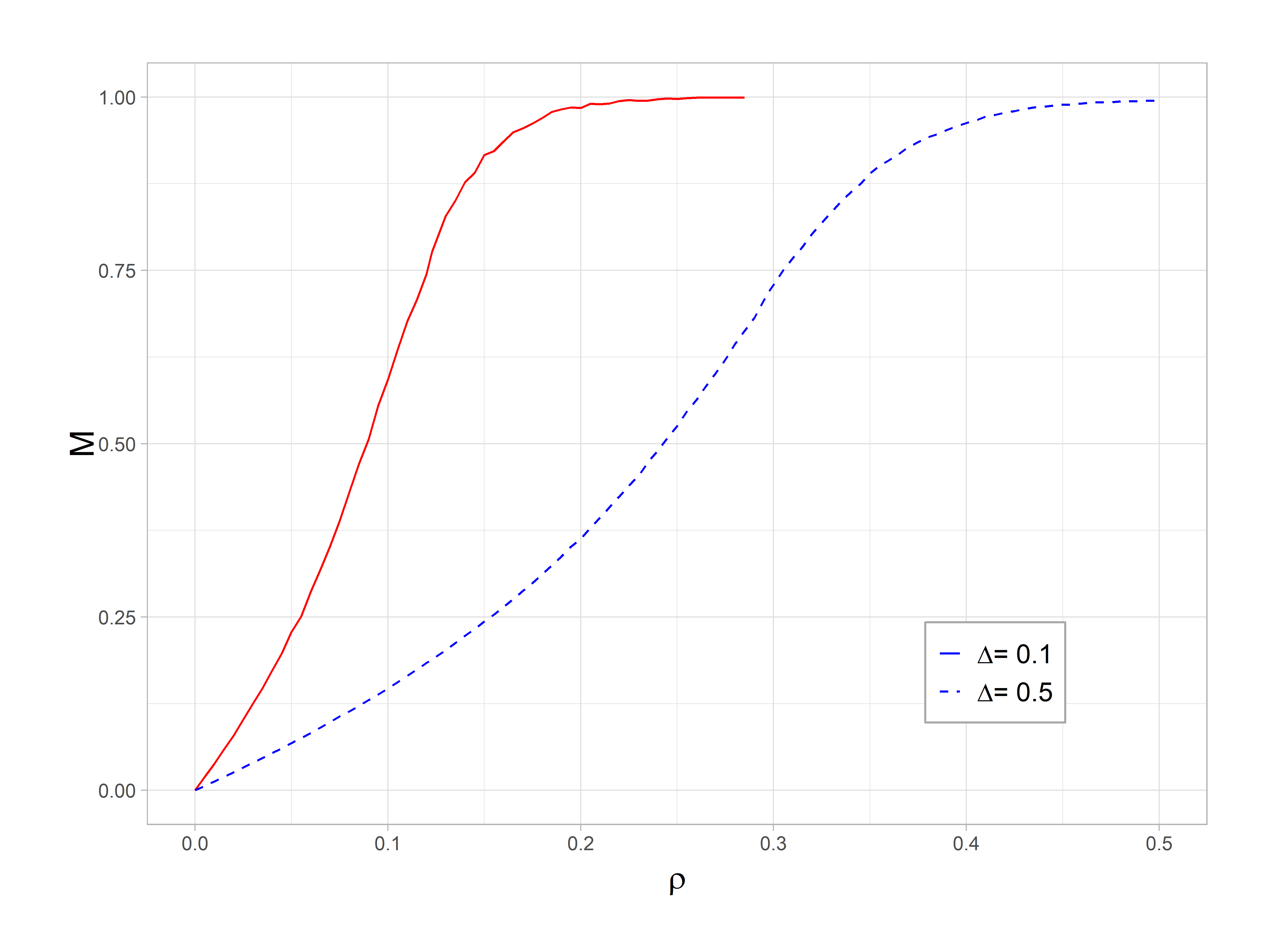}
		\end{center}
		\caption{ The stationary average system state $M$ for   $(r=0.6)$  as a function of the parameter $\rho$ for two different values of $\Delta=0.1$ and $\Delta=0.5$.}
		\label{fig:AA}
	\end{figure}

	In Figure \ref{fig:AA}, we study the behavior of the stationary order parameter $M$ as a function of $\rho$ for different values of $\Delta=0.2, ~0.5$ when $r=0.6$. The order parameter increases with $\rho$. However, the growth rate is higher for lower delta values. The continuous phase transition, from the state $s=0$ to the absorbing state $s=1$ take place at the specifics points $\rho_c$  which depend of  $r$ and $\Delta$.f $\Delta=0.1$ and $\Delta=0.5$.
	Notice that higher values of $\Delta$ implies higher values of  $\rho_{c}$.

\begin{figure}[htb]
		\begin{center}
		\includegraphics[height=8cm, width=8cm]{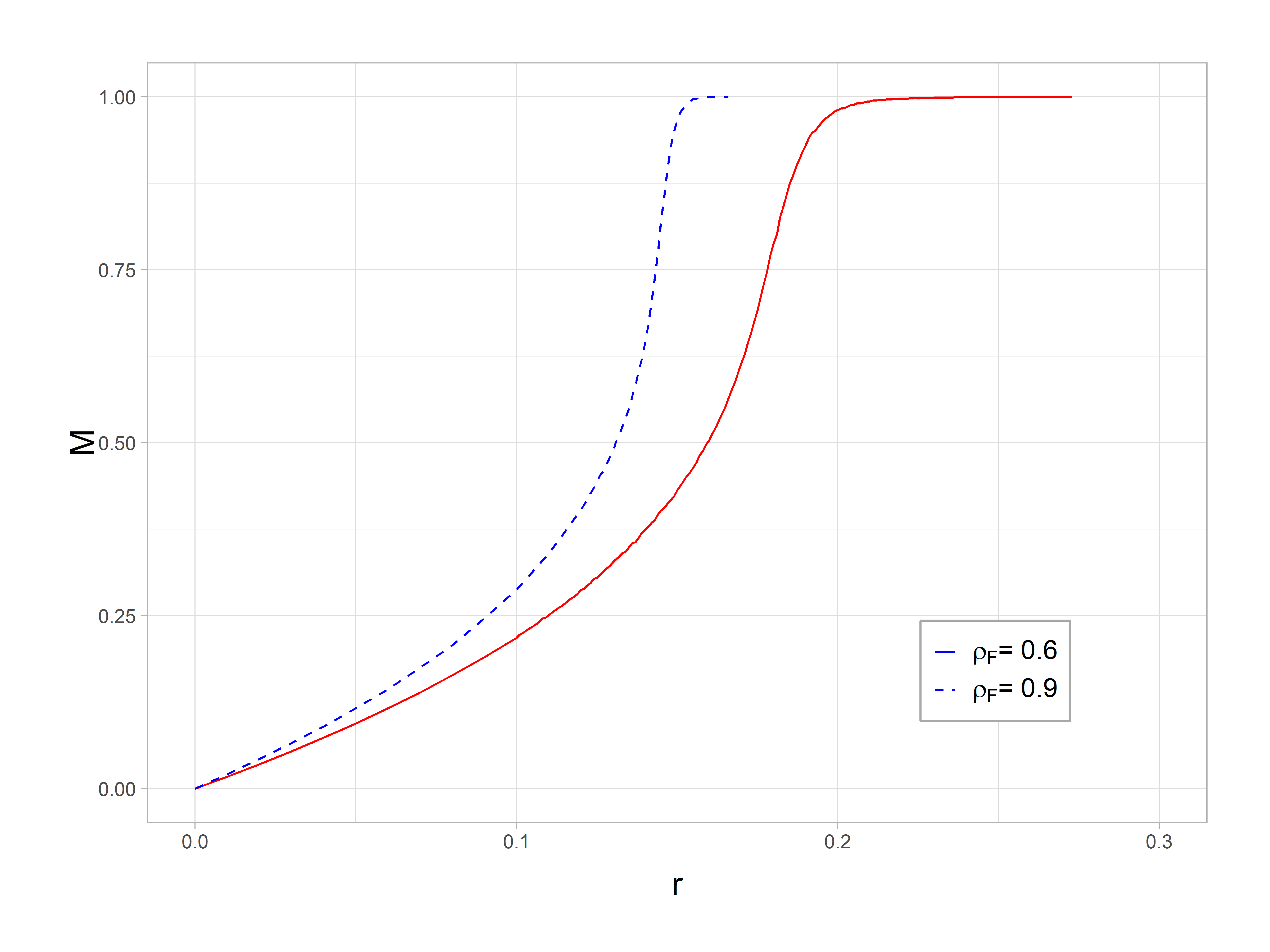}
		\end{center}
		\caption{ Order parameter $M$ for $(\Delta=0.5)$  as a function of the parameter $r$ for different values of $\rho=0.6$ and $\rho=0.9$.}
		\label{fig:B}
	\end{figure}

	In the second projection analyzed, we computed the order parameter as function of $r$ for different values of $\rho$ and $\Delta$. The phase transitions of $M$ take 
	place 
	at the specifics points $r_c$  which depend of $\rho$ and $\Delta$ as we see in  Figure \ref{fig:B}. The higher values of  $\rho$ the higher values of  $r_c$.

\begin{figure}[htb]
		\begin{center}
						\includegraphics[height=6cm, width=8cm]{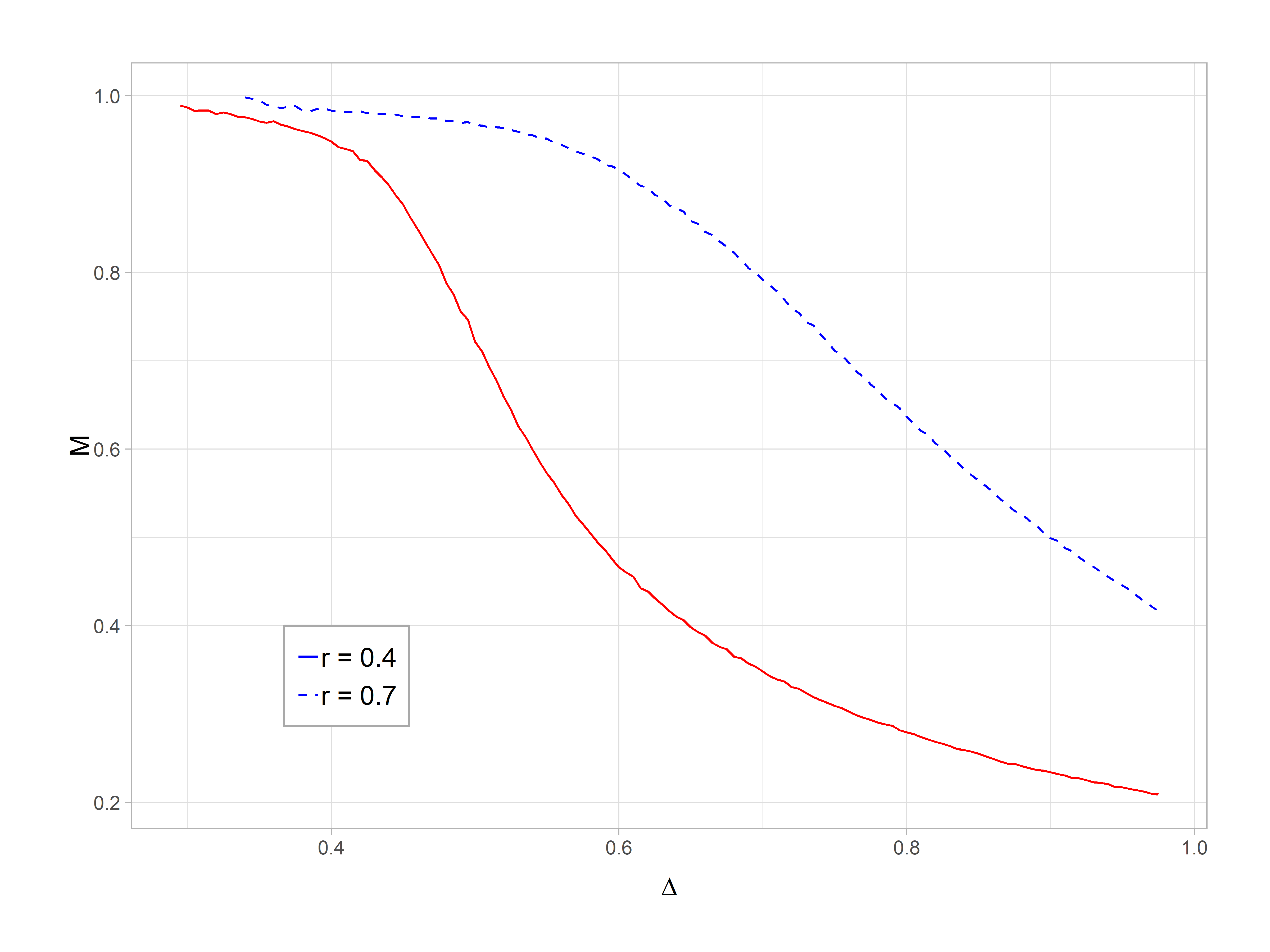}
						
		\end{center}
		\caption{The order parameter  $M$ for $(\rho=0.4)$  as a function of the parameter $\Delta$ for two different values of $r=0.4$ and $r=0.7$.}
		\label{fig:C}
	\end{figure}

	In the third projection we analyze the behavior of the order parameter $M$ as function of $\Delta$ for a fixed value of $\rho=0.4$. In the figure 
	\ref{fig:C} we plot the stationary order parameter as a function of $\Delta$ for two different values of $r=0.4$ and $r=0.7$. The values of $M$ start from 1 and decrease continuously as $\Delta$ increase. The critical point $\Delta_c$  depends on $\rho$ and $r$.

\begin{figure}[htb]
		\begin{center}
			\includegraphics[height=10cm,width=9cm]{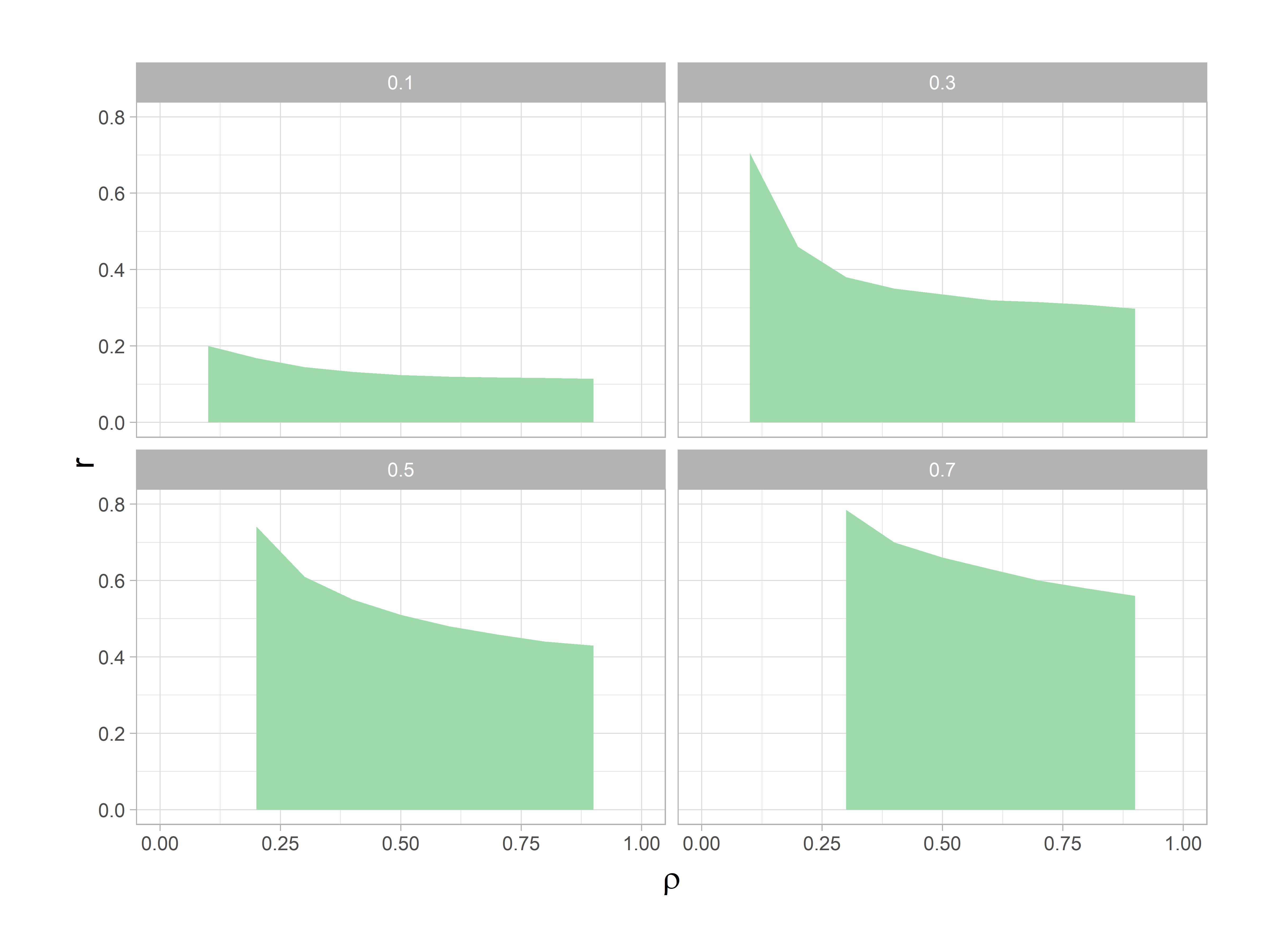}
		\end{center}
		\caption{Phase Diagrams $r \emph{---} \rho$. each panel correspond to different values of $\Delta =(0.1,0,3, 0.5, 0.7)$.}
		\label{fig:D}
	\end{figure}

	In Figure \ref{fig:D} we study the phase diagram $r\emph{---} \rho$ for different values of $\Delta$. The dark color region represents the state $s=0$ while the white region represents the state of the master node $s=1$. Now it is clear the influence of the intrinsic individual feature capture by the control parameter $\Delta$. When it is small, the master state dominates the phase diagram and its influence decreases as the control parameter value increases. 
	
	\section{Discussion}
	\label{sec:discussion}
	
	The state $s=0$ may represent a state in which there is a dominance of agents that are contrarians, egoist, defective, dishonest, corrupt, etc. On the other hand, $s=1$  may represent the opposite, with agents that favor support, faith, cooperativeness, honesty which are values represented in the master node. The parameter $\Delta>0$ is the propensity of an individual to be against the master and it is an intrinsic individual property. Despite it, due to the interactions, opinions or beliefs may spread during the dynamics involve connected peers. In the model we have a stochastic dynamic described by the transition rules in table \ref{tab1}. In the level of network, due peer pressure individuals may change sates (opinions) if there some majority around. However, this is not trivial, since there is some competition between master and contrarians peers.
	
	The master node has an important role in the model. One of the most ubiquitous mechanisms guiding people in deciding who or what to follow is the reinforcement. The master node represents an idea or value and its role is to increase order or promote consensus, although in a biased way. There are two parameters regarding the master node. First is the number of individuals connected to it, which is measured by the density $\rho$. It is not a necessary condition to all individuals be connected with the master to induce a transition to an ordered phase. Second, the power of influence over an individual is in general limited by a quantity $0<r < 1$. Of course, when $r$ is small, it means that the prestigiousness of what the master represents is either in crisis or threatened. However, with more connections, the master node may still continue to exert relevant influence if the product $r \rho$ is above a certain level. In the model we treat each parameter as an independent one, but in the Figure \ref{fig:D} we see a critical line relating both parameters. The master node promotes positive feedback, which creates clusters of individuals with state $s=1$. These clusters influence change of state in a few individuals that are against the orientation, in other words, they promote change from state $s=0$ to state $s=1$. Without action of the master, the interaction at the level of the network evolve to two absorbing states guided by the initial conditions. This is a consequence of the rules of majority underlining the model.

	We notice that our model is endowed with quenched disorder on the connection between the master and some individuals in the network. In real systems both states and network coevolve. Here the quenched disorder in the master node and the network is important. Not all individuals may receive direct information and yet a consensus may emerge. People may be looking for a partners that resemble more their  own properties or beliefs. The quenched disorder in links is one of the model limitations. It was meant to keep the model simple, however, some individuals may abandon or break with collective ideals or beliefs, while others may establish connection with the master. In this case, the density $\rho$  may fluctuate. Also, we address a unidimensional network to mimic society, while social networks are more complex. Those effects have not been studied here.
	
	Phase transition appears by changing parameters. It would be interesting to changes interaction rules that can lead the density $\rho$  to $\rho_c$. In this case, the relationship between master and the individuals may be self-organized around the critical point or critical line. What we know in real systems is that it is very hard to reach the ordered state $(M=1)$. This means that it is not easy to move in the space parameter. Morever, once the parameters are set around critical points, the relaxation time to the systems get order may be very long, with large population $L$. Somehow social systems selforganized in a hierarchical topology similar figure \ref{fig:model}. But the links still evolving and maybe it will be approach critical values.
	
	Although the rules of interaction are simple, we uncover a rich scenario of collective behaviors. The major evidence is given by the phase diagram presented in the previous section. The model analyzed here shows the existence of critical values in several parameters. 	Figures (\ref{fig:AA} to \ref{fig:C}) are two-dimensional projections of the order parameter in a direction of each one control parameter. In Figure \ref{fig:D}) we try to illustrate the volume of the phase space which the coordinates are the control parameter $ \Delta, ~r$ and $\rho$. In the inner part of this volume the order parameter reach its maximum value $M=1$. The shape in this figure is just illustrative. What calls our attention is the properties of the surface of this volume: it separates the synchronized phase where every elements enter in the absorbing state $s=1$ and the phase where there is a mixture $0<M<1$. This idea is corroborated by Figure \ref{fig:C}. We fixed a plan by choosing specific values of $\Delta$. After, we varied $\rho$ and $r$ and we found a critical line splitting two phases. This imply the existence of a critical surface in the 3D phase diagram.
	
	The critical exponents  $\Lambda_c$  along the manifold surface likely are non universals since they may exhibit a continuous dependence of the exponents with the critical control parameters  $\Lambda_c (\Delta,r,\rho)$. This phenomena is represented by small lines leaving the critical surface (see Figure \ref{fig:C}) just to give some ideal of a richness of the phase transition occurring in this system.


	\begin{figure}[htb]
		\begin{center}
			\includegraphics[height=5cm, width=8cm]{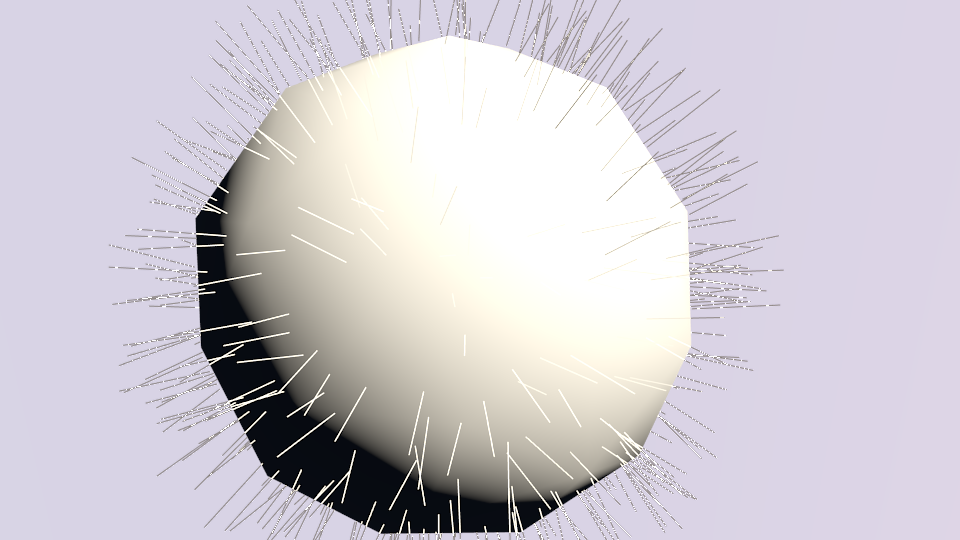}
		\end{center}
		\caption{Illustrative view of a critical surface. The lines leaving the critical surface illustrate dependence of the critical exponent with parameters $\Delta$, $1-\rho$ and $r$.}
		\label{fig:manifold}
	\end{figure}

	\section{Conclusion}
	\label{sec:conclusion}
	
	In the present work we proposed a stochastic quenched disorder model to investigate the power of a master node over a system formed by $L$ elements disposed in a ring network with first neighbor interaction. This system is similar to situations we find in schools, armies, churches and State where teachers, priests and the government may incapsulate the masters’ orientation. Specific discipline, beliefs and citizenship are the goals to be accomplished. In all those cases, we want to know in what extent the all system align to the master orientation for several different scenarios described by the control parameters.
	
	Through Monte Carlo simulations we observe that in the absence of $\rho=0$) the model has two absorbing states. One of these states is formed when all individuals are in the reference state $s=1$ or against $s=0$. When the master is allowed to connect with individuals or other elements in the network $\rho>0$) , the model presents a second-order non equilibrium phase transition. The order parameter (fraction of individuals in state $s=1$) reaches its maximum value $(M=1)$ continuously.
	
	At this point, static and dynamic critical exponents are not universal, i.e. they depend on the initial density of master link ρ which is kept fixed at each instant of time. However, such exponents do not satisfy the generalized hyperscaling relation \cite{hexner2015hyperuniformity}, that is characteristic of systems with infinite absorbing states.
	
	In summary, by mean of a simple model to quantify the level of agreement of a society partially receiving information from a master (representing laws, moral, beliefs) with a relative influence $r$, may lead to a synchronization with a critical number of links $\rho_c$ , $r_c$ or $\Delta_c$. This study may be extend in many different ways. There are many open questions derived from this model, such as the shape of the critical surface in the phase space; the dependence of the critical exponent with the control parameter; the role of the network topology for consensus,and if there exist some analytical solution just to mention a few.
	
	This study may be extend in many different ways. There are many open question derived from this model as the shape of the critical surface in the phase space; the dependence of the critical exponent with the control parameter; the role of the network topology for consensus,and if there exist some analytical solution just to mention a few.
	
	All the questions addressed go beyond the parametrization studied here. Moreover, due the map between the Master Equation \cite{kanpen} and the Schr\"odinger equation \cite{legeto,alef0} it is possible connect a stochastic one-dimensional model in a quantum chain model. In other words, this model can be studied in many different contexts, such as: Quantum Dimers \cite{alef1}, Quantum Quenched Field Theory \cite{alef2}  and Quantum Integrability \cite{alef6,alef7}. Besides that, the operator stocastic interaction corresponds  to $SU(2)\times U(1)$ algebra, i.e; the same simmetry of  Exotic Kondo Model \cite{kondo}. However, the "$\aleph$" stochastic operator  has  special (on-off) terms  of three and four body.
	
	By virtue of these connections we are compelled to roughly say that for every set of moral axioms there exists an equivalent Quantum Field Theory, at least within a platonic space (Model Theory \cite{goodidea}). We hope that this simple model may shed light to improve our understanding on a class of problem that is relevant from view point of science and society.

	\section{Acknowledgments}
	
	The authors are partially supported by CAPES grants. Besides that, the authors are grateful to Alexandre Diehl and Nilton Manuel E. do Rosario for fruitful discussions on Moral Theory.
	
	\bibliography{masternodenetwork}
	
\end{document}